\documentclass[conference]{IEEEtran}
\IEEEoverridecommandlockouts
\usepackage{cite}
\usepackage{amsmath,amssymb,amsfonts}
\usepackage{algorithmic}
\usepackage{graphicx}
\usepackage{textcomp}
\usepackage{xcolor}
\def\BibTeX{{\rm B\kern-.05em{\sc i\kern-.025em b}\kern-.08em
    T\kern-.1667em\lower.7ex\hbox{E}\kern-.125emX}}
\usepackage{times}
\usepackage{epsfig}
\usepackage{graphicx}
\usepackage{amsmath}
\usepackage{epstopdf}
\usepackage{array}
\usepackage{listings}
\usepackage{color}
\usepackage{amsfonts}
\usepackage{amssymb}
\usepackage{subfig}

\newcolumntype{L}[1]{>{\raggedright\let\newline\\\arraybackslash\hspace{0pt}}m{#1}}
\newcolumntype{C}[1]{>{\centering\let\newline\\\arraybackslash\hspace{0pt}}m{#1}}
\newcolumntype{R}[1]{>{\raggedleft\let\newline\\\arraybackslash\hspace{0pt}}m{#1}}
    
\begin{document}

\title{A Light-Weight Authentication Scheme for Air Force Internet of Things}

\author{\IEEEauthorblockN{Xi Hang Cao and Xiaojiang Du}
\IEEEauthorblockA{\textit{Department of Computer \& Information Sciences} \\
\textit{Temple University}\\
Philadelphia, PA, USA \\
\{xi.hang.cao, dux\}@temple.edu}
\and
\IEEEauthorblockN{E. Paul Ratazzi}
\IEEEauthorblockA{\textit{Air Force Research Laboratory Information Directorate}\\
Rome, NY, USA \\
edward.ratazzi@us.af.mil}
}

\IEEEoverridecommandlockouts
\makeatletter\def\@IEEEpubidpullup{2\baselineskip}\makeatother
\IEEEpubid{\parbox{\columnwidth}{\footnotesize DISTRIBUTION A.   Approved for public release: distribution unlimited. Case Number 88ABW-2018- 3936  20180807 
}
\hspace{\columnsep}\makebox[\columnwidth]{}}

\maketitle

\begin{abstract}
Internet of Things (IoT) is ubiquitous because of its broad applications and the advance in communication technologies. The capabilities of IoT also enable its important role in homeland security and tactical missions, including Reconnaissance, Intelligence, Surveillance, and Target Acquisition (RISTA). IoT security becomes the most critical issue before its extensive use in military operations. While the majority of research focuses on \textit{smart} IoT devices, treatments for legacy \textit{dumb} network-ready devices are lacking; moreover, IoT devices deployed in a hostile environment are often required to be \textit{dumb} due to the strict hardware constraints, making them highly vulnerable to cyber attacks. To mitigate the problem, we propose a light-weight authentication scheme for \textit{dumb} IoT devices, in a case study of the UAV-sensor collaborative RISTA missions. Our scheme utilizes the covert channels in the physical layer for authentications and does not request conventional key deployments, key generations which may cause security risks and large overhead that a \textit{dumb} sensor cannot afford. Our scheme operates on the physical layer, and thus it is highly portable and generalizable to most commercial and military communication protocols. We demonstrate the viability of our scheme by building a prototype system and conducting experiments to emulate the behaviors of UAVs and sensors in real scenarios.    
\end{abstract}


\section{Introduction}

\subsection{The Smart Things and the Dumb Things}

Internet of Things (IoT) provides us with unprecedented opportunities \cite{gubbi2013internet}, it also poses challenges, such as privacy and security \cite{miorandi2012internet}. In general, ``things" in IoT refers to devices that are \textit{smart}, meaning that they are capable of computing, sensing, actuating, and being connected via the Internet; these smart things can be found across our lives, from traditional IT orientated devices \cite{du2006adaptive}, such as smartphones and laptop computers, to more living orientated devices, such as smart lights, smart appliances, and electronic personal assistance. In a more broad sense, ``things'' in IoT may also include sensors and actuators that equip with extremely low computing capabilities \cite{atzori2010internet, du2005designing, mandala2008load}, and we call these devices \textit{dumb}. These dumb things may be legacy devices in which the electronics and designs are based on an older generation of technologies, and these dumb things may be mission-specific devices in which trade-offs are made between functionalities and special mission requirements, such as low-power consumption, tiny size, etc. In addition to the traditional IoT security challenges in the scenarios of only smart things, the security challenges in the scenarios of both smart things and dumb things are considerably more critical and difficult to overcome, due to that the lack of computational capability in the dumb things makes conventional security strategies (e.g., key generation \cite{du2008security, du2004qos}, key management \cite{du2009routing, xiao2007survey, xiao2007internet, du2007effective}, etc.) inapplicable. In this study, we aim to tackle the challenge in the latter scenario in which we propose a light-weight authentication strategy for dumb authenticators.

\subsection{IoT in the Air Force and Our Use Case}

As the technologies encapsulated in the realm of the IoT have become mature, the US military, especially the Air Force (AF), has identified the tremendous application potentials of the IoT in operations and missions. The paradigm of IoT has been adopted in different areas. In addition to the IoT applications in the commercial and civilian domains, the IoT adoption in the AF also spans its applications in the battlefield and tactical environments. 

\begin{figure}[t]
\centering
\includegraphics[width=7cm]{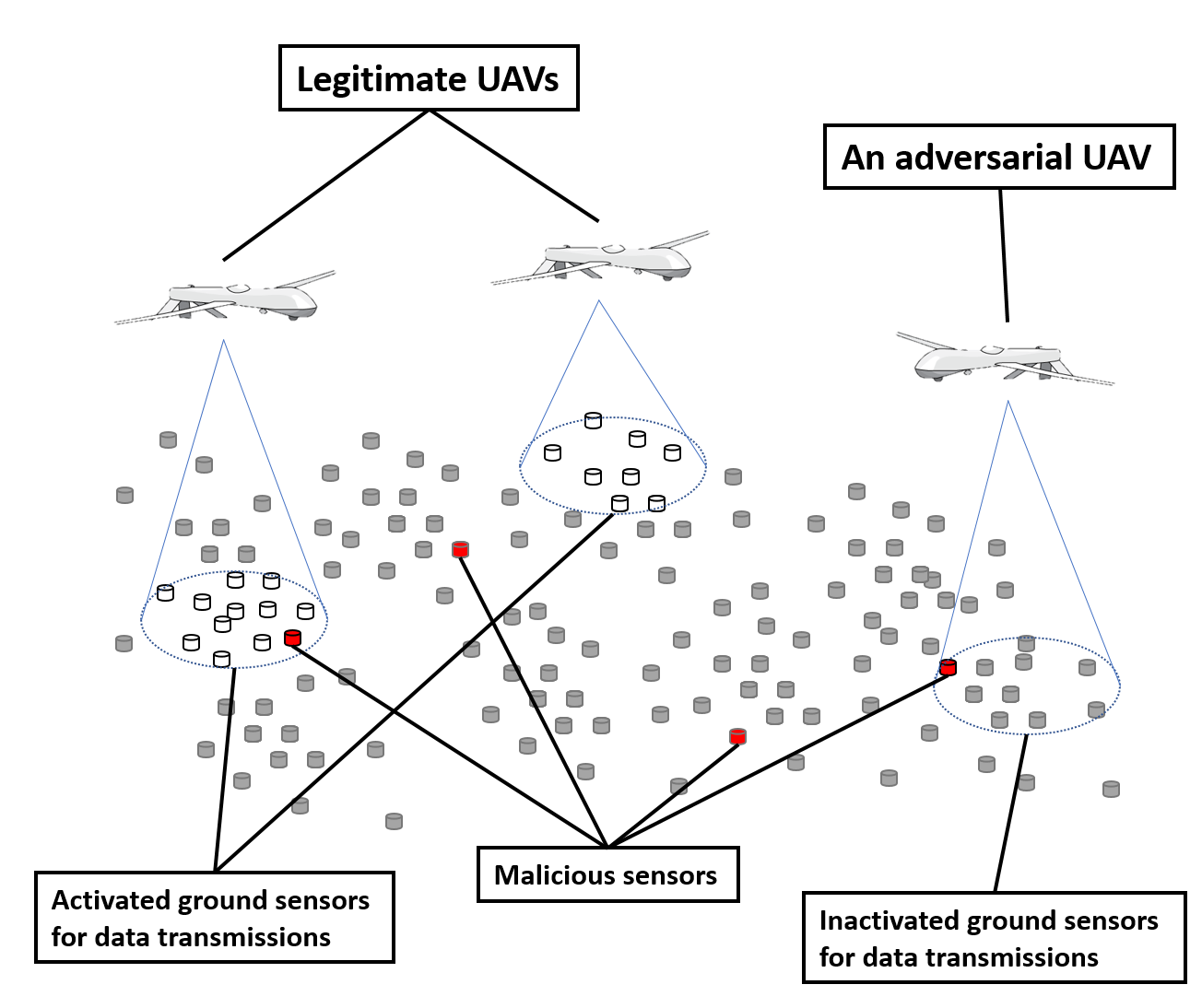}
\caption{A demonstration of the air-ground collaborative surveillance operation, in which UAVs are the platforms to collect and process the data obtained from the ground sensors. In a battlefield or hostile environment, adversarial UAVs and sensors are expected. Therefore, an effective authentication scheme is crucial for mission success.}
\label{fig:use-scenario}
\end{figure}

A typical use case in the RISTA missions is shown in Fig.~\ref{fig:use-scenario} \cite{ho2011performance} where a general architecture of UAV-sensor network system is demonstrated. Usually, this use case appears in the air-ground cooperative surveillance operations \cite{grocholsky2006cooperative}. A collection of small, low-powered, low-memory sensors reside at distributed geo-locations. These sensors operate in two modes: a sensing mode and a transmission mode. In the sensing mode (low energy consumption), sensors make environmental measurements and write the data into memory; in the transmission mode (high energy consumption), the sensors activate the link to the UAV and transmit the stored data. The transmission mode is activated by beacons when a UAV flies above them, and the transmission is carried on as the UAVs maintain an effective communication distance. Because of this data collecting process, sensors can preserve energy by avoiding constant data transmission which may cause high energy consumptions; adequate frequent data transmissions are required to prevent data loss due to memory cycling. The UAV performs as a data transmission relay or a data fusion platform to overcome communication and data transmission problems caused by distance and blockages.  

\subsection{Contributions}

We propose an approach which is based on covert channels in the physical layer. Specifically, we encode secret patterns within the physical layer controllable parameters, including, transmission power level, transmission frequency, and beacon intervals. These physical layer covert channels are incredibly portable and generalizable to various commercial and military communication standards and are simple and effective enough that will not cause computational overhead and thus suitable for \textit{dumb} authenticators. More details of our approach can be found in Section~\ref{sec:proposed-approach}.

\section{Related Works}\label{sec:related-work}

Our authentication scheme combines the advantages of two ideas: physical layer fingerprints and covert channels. In this section, we briefly review some state-of-the-art works in this two research areas.

\subsection{Related works in Authentications Using Physical Layer Fingerprints}

It has been demonstrated by many previous studies that by utilizing the uniqueness of the physical layer features, RF/wireless devices can be identified with high accuracy; for example, RFID device identifications \cite{danev2009physical} and transmitter identifications \cite{liu2008specific}. Moreover, the physical layer fingerprint (or specifically, radio frequency fingerprint, RFF) has been used in wireless security \cite{xu2016device}, including, authentication \cite{xiao2008using} and intrusion detection \cite{hall2005radio}. It is notable that some of the previous works focused on devices with some specific components, so those features are not generalizable to many applications. Usually, features are extracted from time domain, frequency domain, and wavelet domain. Time domain features (also called radiometric) include amplitude, phase, and other more complex feature that usually by comparing the obtained waveform to a reference/ideal waveform, for example, clock skew variation \cite{kohno2005remote}, transient properties \cite{hall2006detection}, waveform accuracy \cite{gerdes2006device}, etc. Frequency domain features are usually obtained by applying Fourier transform to the waveform. For enhancing the identification power, combinations of the feature are used.  

\subsection{Authentications Using Covert Channels}

There has been research about covert channels in wireless communications, in both design and detection. Usually, a covert channel is any communication channel that can be exploited by a process to transfer information in such a way that breaks a system's security policy \cite{lampson1973note}; therefore, a covert channel poses challenges in information security. For more efficient detections, one has to explore the possible covert channel, so there had been many covert channel designs proposed. Many of them can be categorized into either storage covert channel \cite{singh2015establishment, denney2016novel} or timing covert channels \cite{zhao2015wlan}. In a storage covert channel, the covert message is encoded into a particular portion of the legitimate traffic; in a timing covert channel, a covert message is inserted by manipulating the typical time-based properties of the system, such as CPU time and inter-packet delay \cite{heda2015covert}. Some recently proposed covert channels have explored other perspectives of the communication system, for example, power \cite{wang2017classifier}, radio waveform \cite{guri2016usbee}, and modulation schemes \cite{cao2018wireless}. Our approach for UAV authentication is similar to covert channel communication, in the sense that the information transmitted is ``hidden" within a medium. In our approach, the information of the secret pattern is hidden within the transmission power, frequency/channel, and beacon interval, which we call physical layer covert channels. 

\section{Models}\label{sec:models-and-assumptions}

\subsection{System Model}
\label{subsec:system-model}

\begin{figure}[t]
\centering
\includegraphics[width=6cm]{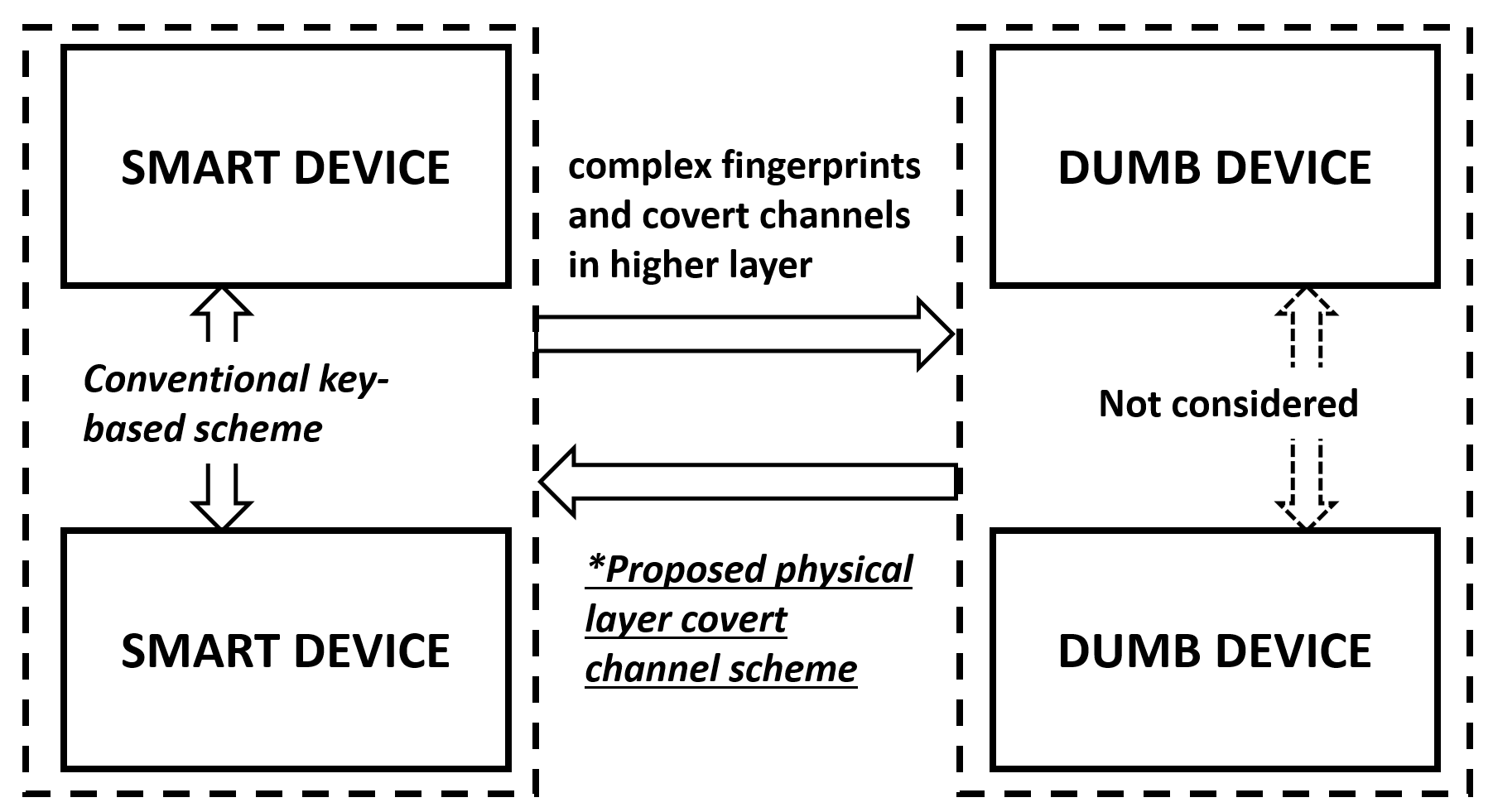}
\caption{A system model schematic depicts the various types of authentication schemes to consider}
\label{fig:system-model-block-diagram}
\end{figure}

In our system, we consider two types of devices. One type is called smart device, which is capable for rigorous tasks, such as computing, communicating, and actuating; in our use scenario, this type of device refers to the UAVs for collecting data from the field. Another type is called dumb device, whose capability is limited by electronic design and mission requirements. For power sustainability, in the majority of the time, dumb devices operate in a low-power consumption mode for environmental sensing and data recording. Once activated, they will operate in the data transmission mode, in which the dumb devices transmit their stored data to an authorized party and clear their memory for the coming record cycle. The schematic of the possible mutual authentication is shown in Fig.~\ref{fig:system-model-block-diagram}.
When the authenticator is a smart device, we have the flexibility to adopt conventional IoT security strategies \cite{conti2018internet}. The challenges lie in the cases that the authenticator is a dumb device which is not compatible with conventional strategies (e.g., key-based schemes). Our proposed approach focuses on the case in which a dumb authenticator authenticates a smart device. The authentication between two dumb devices is not considered in this study.


\subsection{Threat Model}\label{subsec:threat-model}

We consider a smart adversarial device (UAV), which is trying to communicate with the dumb devices (sensors) and request the stored sensing data (the UAV on the right of Fig.~\ref{fig:use-scenario}). The adversarial UAV sends beacons to the sensors below it and try to activate their transmission mode; however, because of our proposed physical layer convert channel authentication scheme, the adversarial UAV will not pass the authentication process, and thus, the sensors will refuse to transmit the stored data. Adversarial parties may also place malicious sensors into surveillance terrain to send misleading/confusing data (as shown in Fig.~\ref{fig:use-scenario}). In this case, a legitimate UAV can identify the malicious sensors by recognizing the abnormal data transmission timings and unregistered physical layer fingerprints.   

Beyond the above threats, the threats from traditional Internet \cite{stallings2000network} are also applicable to our used scenario. In the following, we list the threats (their breaches of security goals):
\begin{itemize}
\item Message Replay (Authentication)
\item Impersonation (Authentication)


\item Eavesdrop (Confidentiality)
\item Man-In-The-Middle (Integrity, Confidentiality)
\end{itemize}

\section{Proposed Approach: Authentications based on Physical Layer Covert Channels and Secret Communication Patterns}\label{sec:proposed-approach}

The constraints of hardware and the lack of computational power in dumb devices pose significant challenges in exploiting a wide range of existing effective physical layer fingerprints proposed in  the published literature \cite{kohno2005remote}. Without adequate electronic components, the dumb sensors cannot detect the physical layer signals from other devices; nevertheless, without advanced processing units, the dumb sensor cannot extract the hidden features within the signals. We proposed an authentication scheme which leverages the primitive physical layer parameters in communication and use these parameters as covert channels to transmit secret patterns for dumb authenticators. Specifically, we use the physical layer parameters: \textit{transmission power level (txpower)}, \textit{transmission frequency/channel (freq)}, and \textit{beacon interval (beacon\_int)}. These physical layer parameters are so primitive that even without specific electronic components, the dumb authenticators can still detect them; in addition, these physical parameters can be used as covert channels, such that without knowing them \textit{a priori}, one cannot discover these covert channels entirely and easily; more importantly, we utilize these covert channels to transmit secret patterns which can only be known by authorized parties, and this can tremendously enhance the authentication scheme. 

\subsection{Physical Layer Controllable Parameters}

\paragraph{Transmission Power (Txpower) and Received Signal Strength Intensity (RSSI)}

In wireless communications, transmission power (\textit{txpower}) is the amount of energy used when a transmitter emits the radio wave; the level of \textit{txpower} is usually affected by the magnitude, frequency, and modulation scheme based on a communication protocol. Usually, the \textit{RSSI} of a specific \textit{txpower}, follows an exponentially decaying curve as the distance between the transmitter and receiver increases. Although insignificant, the \textit{RSSI} of a \textit{txpower} can be affected by factors of multipath, shadowing, and path loss \cite{goldsmith2005wireless}. Because of the noisy nature of the \textit{RSSI}, while we use \textit{txpower} as a covert channel, we will use the high and low levels to indicate binary digits (1's and 0's) as the transmitted patterns.   

\paragraph{Wireless Communication Frequency/Channel (Freq)}

A particular frequency spectrum is assigned to a specific wireless communication standard. A spectrum may be subdivided into channels with a center frequency and bandwidth. For example, the 802.11bg standards use the 2.4 GHz band which sub-divided into 14 channels spaced 5 Mhz apart. In military wireless communication standards, one can use the \textit{Link-16} standard \cite{li2013communication}, which utilize the L-band (969 - 1206 MHz). For utilizing frequency/channel as a covert channel, a transmitter switches the transmission frequency/channel within the band when beacons are sent to the sensor.    

\paragraph{Beacon Interval (Beacon\_int)}

In our use scenario, dumb sensors are in sensing mode until they receive beacons from a UAV. After the first beacon is received, the sensors keep learning to the following beacons and time the timings of following beacons. The time difference between any two consecutive beacons is defined as the beacon interval. In our covert channel, the beacon interval is measured in a relative unit. Namely, we do not use absolute time (i.e., 100 milliseconds, 200 milliseconds, etc.). Instead, we use the interval between the first and second beacons as a time unit, and any following beacon intervals are integer multiples of this time unit.

\subsection{Covert Channels Design and Authentication Process}

\begin{figure*}[!ht]%
\centering
\subfloat[][]{\includegraphics[width=7cm]{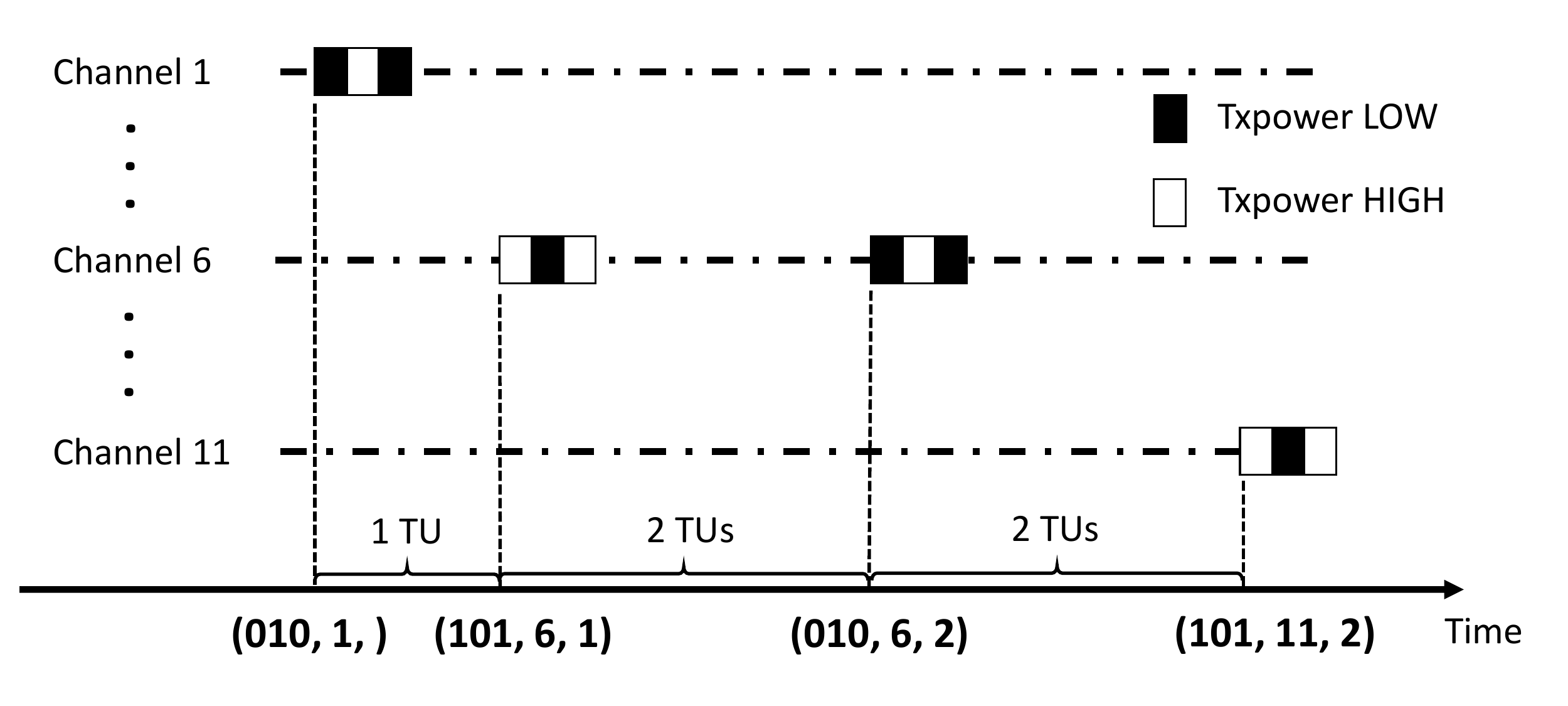}\label{fig:convert-channel-demo-1}}%
\qquad
\subfloat[][]{\includegraphics[width=7cm]{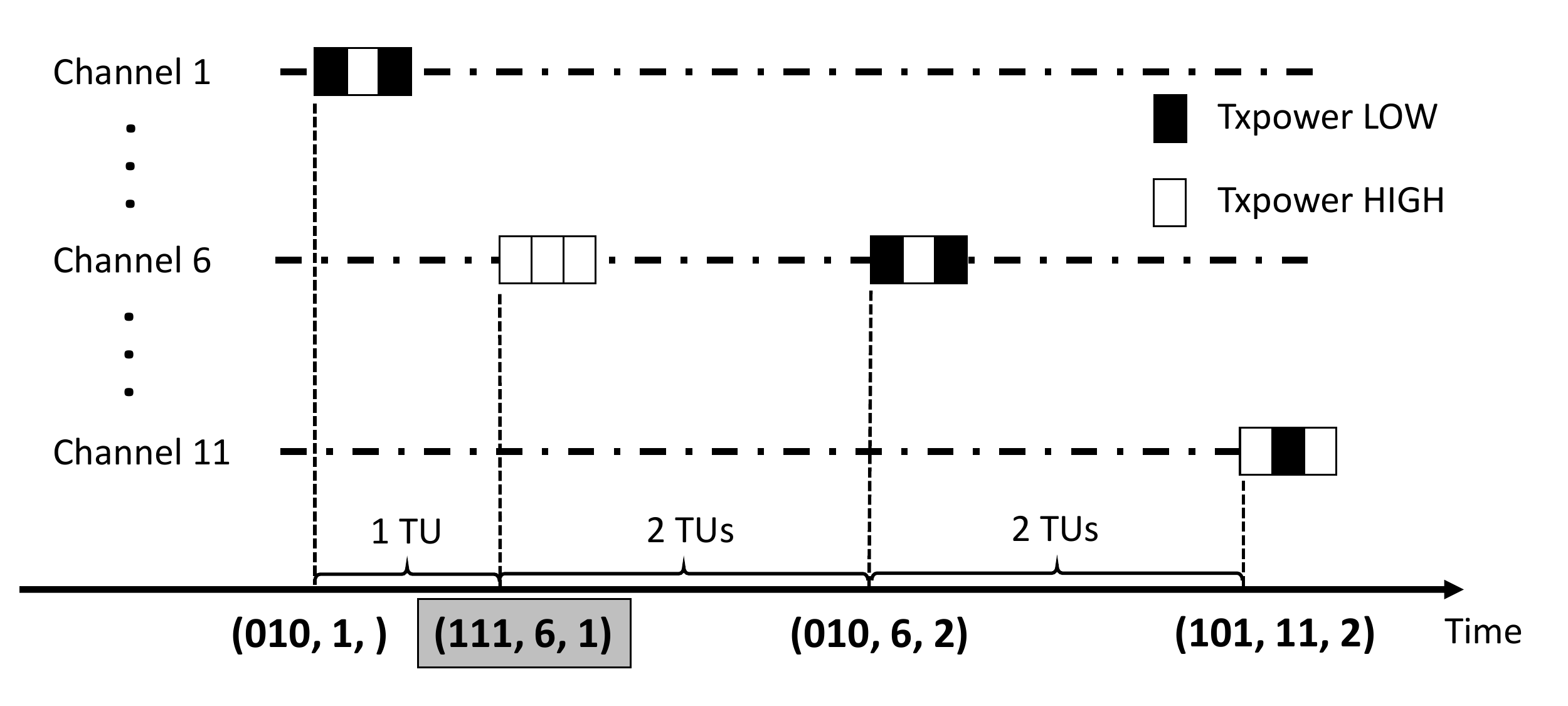}\label{fig:convert-channel-demo-2}}
\qquad
\subfloat[][]{\includegraphics[width=7cm]{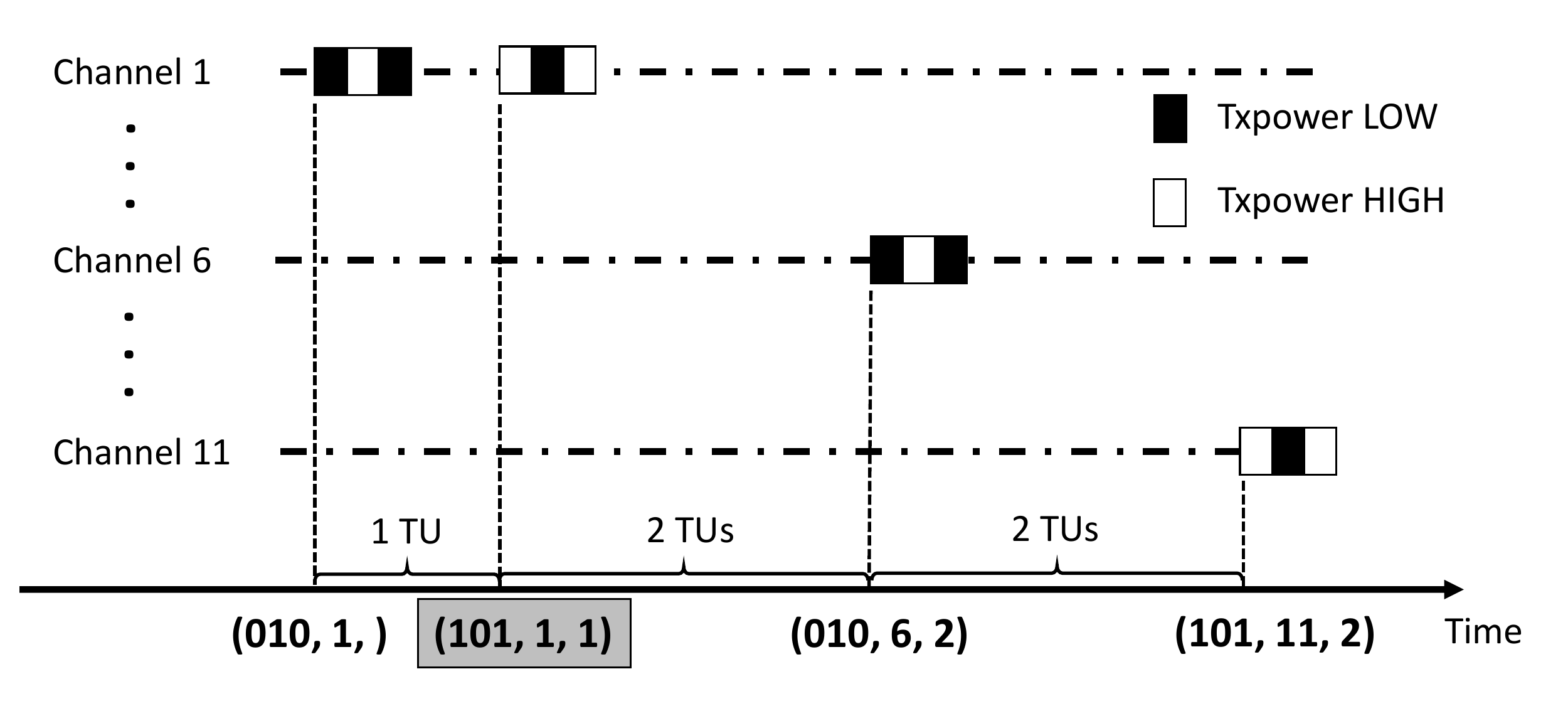}\label{fig:convert-channel-demo-3}}
\qquad
\subfloat[][]{\includegraphics[width=7cm]{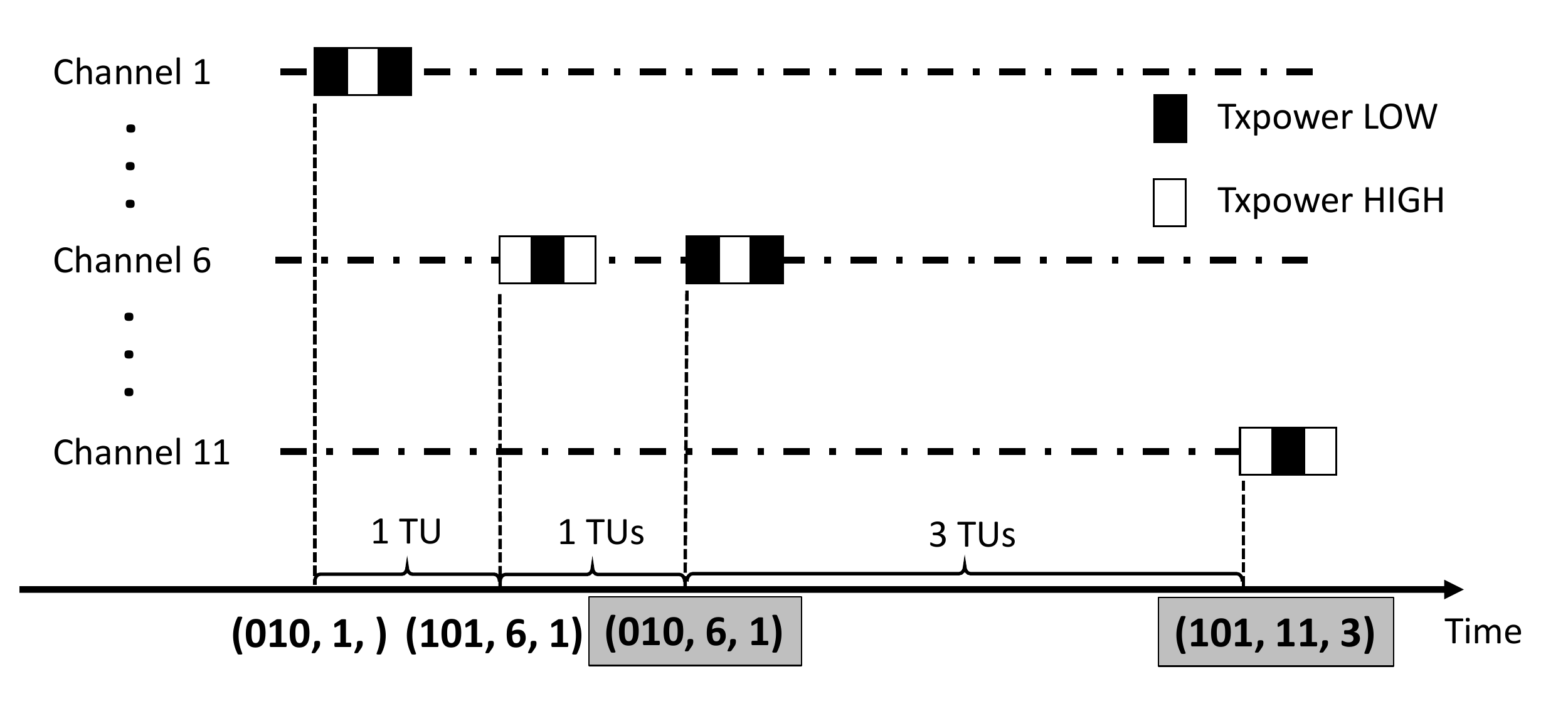}\label{fig:convert-channel-demo-4}}
\caption{An example of a covert channel secret pattern and associated possible incorrect patterns. In the figures, the horizontal direction indicates the time and the vertical direction indicates frequency/channel. (a): the correct sequence of four triplets stored in the memory in the sensors; (b) - (d): incorrect patterns due to the unmatched txpower pattern, channel, and beacon interval, respectively. Unmatched triplets are colored in gray}%
\label{fig:convert-channel-demo}%
\end{figure*}

This section presents our covert channel design using the aforementioned physical layer parameters and shows the authentication process using an example. Before the sensors are deployed, there should be several preset patterns stored or hardcoded in the sensors' memories, such that the sensors can quickly match the desired covert channel patterns versus incorrect patterns.

After a sensor receives the first beacon from a UAV, the sensor will measure the \textit{txpower}, frequency/channels switching, and beacon timings continuously (with an adequately high sampling rate). Based on the observations of these three physical layer parameters, a sequence of triplets are generated and stored: ($txpower\_pattern_0$, $channel_0$, ), ($txpower\_pattern_1$, $channel_1$, 1 TU), ($txpower\_pattern_2$, $channel_2$, $beacon\_int_2$), ($txpower\_pattern_3$, $channel_3$, $beacon\_int_3)$, $ \cdots $. Noting that the subscripts indicate the order of measurements. Upon receiving the first (0-indexed) beacon, the beacon interval is not available, so the third entry (beacon interval) in the first triplet is absent, and because we use the time difference between the first and second beacon as the reference of 1 time unit (TU), the third entry (beacon interval) in the second triplet is always 1 TU. The measurements will carry on until the desired pattern is matched, or terminate if any incorrect triplet appears.

An example of a secret pattern consists of a sequence of four triplets is shown in Fig.~\ref{fig:convert-channel-demo}. The correct pattern is shown in Fig.~\ref{fig:convert-channel-demo-1}, with four triplets: (010, 1, ), (101, 6, 1), (010, 6, 2), (101, 11, 2); Fig.~\ref{fig:convert-channel-demo-2} shows a failed case where in the second beacon, the txpower does not match the correct pattern; Fig.~\ref{fig:convert-channel-demo-3} shows a failed case where the second beacon is not in the correct channel in the communication band; Fig.~\ref{fig:convert-channel-demo-4} shows a failed case where the beacon intervals between the second and third beacon do not match the correct value. 

\subsection{Application Layer Secret Communication Pattern for Authentication}

After a smart device (UAV) passes the physical layer covert channel authentication, a connection between the smart device and the dumb device (sensor) is established. If the dumb device is capable, our authentication scheme has the flexibility to add an extra authentication step before the stored data are transmitted. Specifically, the dumb device requests a secret message in the application layer for authentication. If the application layer message sent by the smart device matches the secret communication pattern, the dumb device will start transmitting the stored data; otherwise, the connection between them will be terminated immediately.

\section{Security Analysis}\label{sec:analysis}

In this section, we analyze the situation that under the proposed authentication scheme, how the possible threats listed in Section.~\ref{subsec:threat-model} are handled:
\begin{itemize}
\item \textit{Message Replay (Authentication):} Messages (data) sent from the sensors are time-stamped or with unique identifiers; therefore, replayed messages will be ignored.  
\item \textit{Impersonation (Authentication):} In order to obtain the credentials in our physical layer covert channel authentication scheme, an adversarial device needs to figure out both the types of covert channel and the covert channel secret messages, which requires $ \mathcal{O}(2^{nL}) $ trials to crack, using a brute force approach, without considering the factors of frequency and beacon interval. The letter $n$ indicates the number of bits in the txpower pattern in each triplet, and the letter $L$ indicates the length of the triplet sequence. Therefore, when the quantity of $nL$ is large (e.g., 128 or above), the secret message is extremely difficult to crack. 

\item \textit{Eavesdrop (Confidentiality):} Theoretically, the threat of eavesdropping can be efficiently handled by encryption; for example, after the transmission is activated between a UAV and a sensor, they can establish a key for symmetric encryption to prevent eavesdropping.     
\item \textit{Man-In-The-Middle (Integrity, Confidentiality):} When there is no attack, a UAV and a ground sensor directly communicate with each other using wireless. For the Man-In-The-Middle (MITM) attack, an adversary needs to relay messages between the UAV and sensor, which  significantly increases the delay of the communication. Hence, the MITM attack may be detected based on the delay of the messages. 
\end{itemize}

\section{Prototype and Performance Evaluation}\label{sec:prototype}

\subsection{Prototype System Setup}

\begin{figure}[ht!]
\centering
\includegraphics[width=7.5cm]{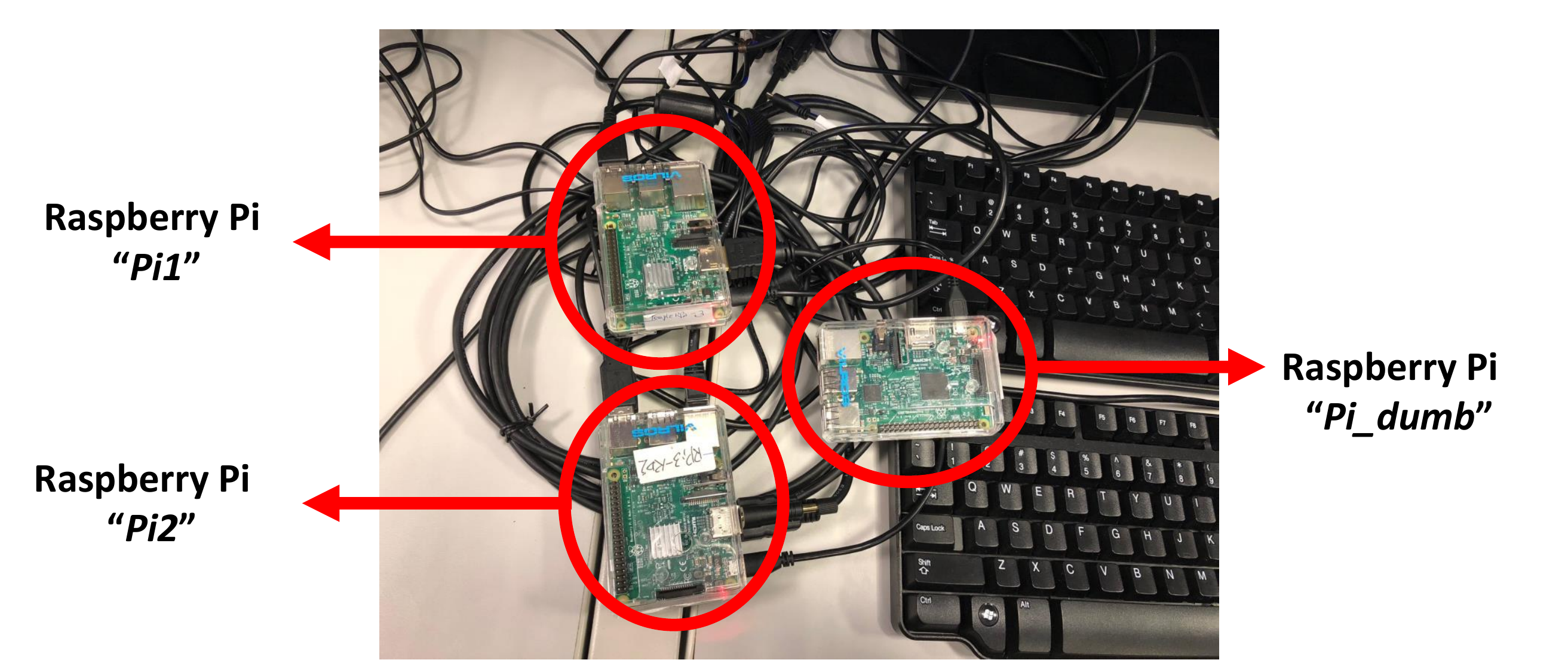}
\caption{A prototype system to emulate the use scenario using three Raspberry Pi boards.}
\label{fig:bench-setup}
\end{figure}

\paragraph{Hardware} We built a prototype system consisted of three Raspberry Pi 3 Model B boards (Fig.~\ref{fig:bench-setup}) to emulate the use scenario in our study. Two of them were configured as WiFi (802.11b) access points, with names \textit{Pi1} and \textit{Pi2}, to emulate the smart devices (UAVs); one of them, with name \textit{Pi\_dumb}, was normally configured, to emulate a dumb device (sensor), which could connect to either of the access points (UAVs).   

\paragraph{Software} The access point functionality was realized by using the \textit{hostapd} tool. To emulate the physical layer covert channel secret patterns, we altered the access point settings, i.e., txpower, channel, and beacon interval, by editing the configuration file \textit{hostapd.conf}. The access point settings can be detected by the \textit{Pi\_dumb}, running the \textit{iw wlan0 scan} command in terminal. To automate the pattern generation and pattern detection process, we created scripts in Python language using the \textit{time} and \textit{subprocess} packages.

\subsection{Extracting Triplets}

Once a Raspberry Pi board is configured as an access point (AP), it will start transmitting beacons to the environment. Any compatible network interface card can detect the received signal strength intensity (RSSI), frequency/channel, beacon interval, and other parameters from the AP. These parameters can be read by using the \textit{scan} command in the \textit{iw} utility or other networking tools.

\begin{figure}[!ht]
\centering
\includegraphics[width=7.5cm]{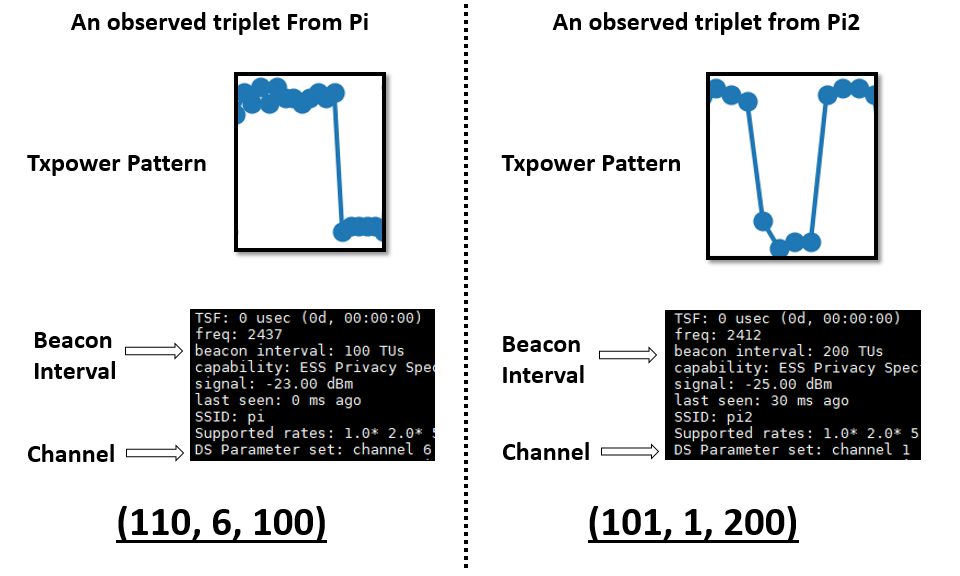}
\caption{Two example triplets to show the difference of the two access point in txpower pattern, frequency/channel, and beacon interval}
\label{fig:triplet-comparison}
\end{figure}

In this experiment, we create a Python script in \textit{Pi\_dumb} and let it automatically and continuously (with a sampling rate of 5 Hz) to detect the wireless parameters from the two APs (\textit{Pi1} and \textit{Pi2}). Fig.~\ref{fig:triplet-comparison} shows two observed triplets generated by \textit{Pi1} and \textit{Pi2}.
In the top penal, we can see the txpower (RSSI) patterns. The txpower patterns generated by \textit{Pi1} and \textit{Pi2} are 110 and 101, respectively. The second panel shows the results returned by the \textit{scan} command in the \textit{iw} utility, and we can obtain frequency/channel, beacon interval and other information from the APs. The triplet extracted from \textit{Pi1} and \textit{Pi2} are (110, 6, 100) and (101, 1, 200), respectively. If we suppose \textit{Pi1} is a legitimate UAV and \textit{Pi2} is an adversarial UAV, then the dumb sensor (\textit{Pi\_dumb}) can easily distinguish them.

\subsection{The Effect of Distance on RSSI}

As the distance increases, the RSSI generally follows an exponentially decaying curve. In this experiment, we measure the RSSI profiles measured by \textit{Pi\_dumb} at different distances as \textit{Pi1} increases the txpower from the minimum (around 7 dBm) to maximum (around 13 dBm). 

\begin{figure}[!ht]%
\centering
\subfloat[indoor]{\includegraphics[width=0.232\textwidth]{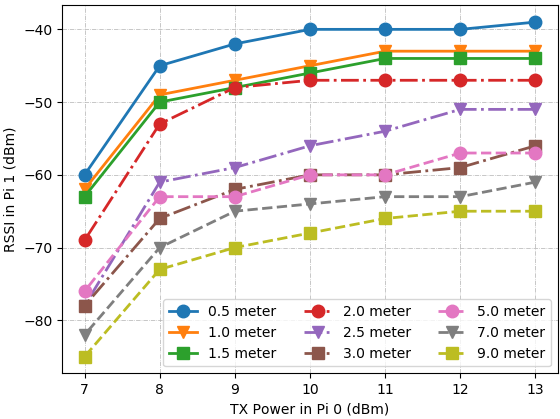}\label{fig:rssi-vs-txpower-range-meter}}%
\quad
\subfloat[outdoor]{\includegraphics[width=0.232\textwidth]{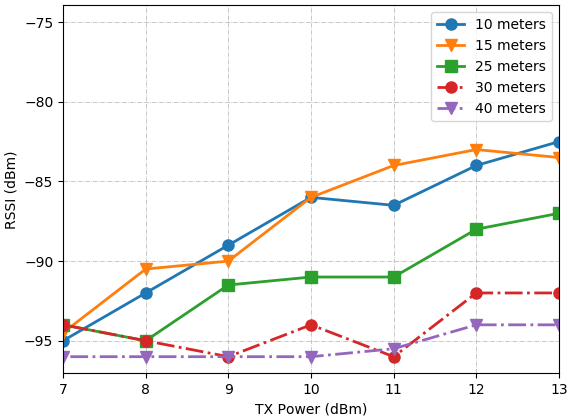}\label{fig:rssi-vs-txpower-long-range-meter}}
\caption{The measured profiles of RSSI as the distance and txpower change in different environments: (a) indoor environment; (b) outdoor environment}%
\label{fig:rssi}%
\end{figure}

\paragraph{High/Low RSSI v.s. High/Low Txpower} The first batch of measurements were made indoor, and the distance was from 0.5 meters to 9.0 meters. The RSSI profiles of the indoor measurements are shown in Fig.\ref{fig:rssi-vs-txpower-range-meter}.

We can notice that when the distance is over 3 meters, the RSSI corresponding to the maximum txpower is lower than the RSSI corresponding to the minimum power at 0.5 meters. Therefore, in our use scenario, as a UAV flies above a sensor, the high/low pattern will be distorted by the changes in distance. To remedy this, we can use a trick: instead of observing the absolute high/low pattern of an RSSI, the sensors observe the steep transitions. 

\paragraph{Maximum Operating Distance} We found that when the distance was beyond 40 meters, \textit{Pi\_dumb} could no longer detect the beacons from \textit{Pi1}. As shown in Fig.~\ref{fig:rssi-vs-txpower-long-range-meter}, when the distance was larger than 30 meters, the RSSI difference between the maximum and the minimum txpower was too small, such that the high/low levels or the transition may not be correctly detected. Therefore, in order for your scheme to operate properly in the prototype system, the Raspberry Pi boards have to be within 30 meters.   
A concern may arise because of that in our use scenario, the distance between a UAV and a sensor may be much larger, and may never be within 30 meters even at their closest positions. The fact is that this maximum operating distance may vary based on the used wireless communication standard and hardware quality. In our use scenario, military standards may be used, and the txpower of a UAV transmitter will be much higher than the txpower of a Raspberry Pi transmitter; in addition, the antenna sensitivity in a military-grade sensor will be higher than the sensitivity of a Raspberry Pi receiver. Therefore, it is reasonable to believe that the maximum operating distance in our use scenario will be much larger than 30 meters.

\subsection{Application Layer Secret Messages}

After the connection between a dumb and a smart device is established, an additional authentication step based on application layer message may be included for an extra level of security. If the secret message sent by the UAV is correct (i.e., 1234567890 in the demonstrated case), data transmission starts; otherwise, the connection will terminate immediately.

\section{Conclusions}\label{sec:discussion-and-summary}
In this paper, we designed a light-weight authentication scheme based on physical layer covert channels. Our scheme tackles the challenges in the IoT scenarios where both smart and dumb devices are present, for example, the air-ground collaborative reconnaissance, intelligence, surveillance, and target acquisition (RISTA) operations with capable UAVs and mission-specific (legacy or dumb) sensors. We conducted a rigorous security analysis and built a prototype system to demonstrate that the proposed scheme is effective, portable, and generalizable to real scenarios.

\bibliographystyle{IEEEtran}
\bibliography{reference}

\end{document}